# Nesting properties and anomalous band effect in MgB$_2$

Izumi HASE[*] and Kunihiko YAMAJI

*Nanoelectronics Research Institute, AIST , Tsukuba, 305-8568*

First principle FLAPW band calculations of the new superconductor MgB$_2$ were performed and the polarization function $\Pi_{12}(\mathbf{Q})$ between the two π-bands mainly formed of boron p$_z$-orbital was calculated. We found that $\Pi_{12}(\mathbf{Q})$ is substantially enhanced around $\mathbf{Q}=(0,0,\pi/c)$, which supports the two-band mechanism of superconductivity for MgB$_2$. $\Pi_{12}(\mathbf{Q})$ peaks at $Q_z \sim 0.3(2\pi/c)$ and $Q_z \sim 0.5(2\pi/c)$. These two peaks are related to the nesting of these Fermi surfaces, but significantly deviates from the position expected from the simplest tight-binding bands for the π-bands. From the calculations for different lattice parameters, we have found significant dependences on the isotopic species of B and on the pressure effect of the polarization function in accordance with the respective changes of $T_c$ in the above-mentioned framework.

---

[*] mailto: i.hase@aist.go.jp

The prominent $T_c$ of MgB$_2$ among various AlB$_2$-type diborides[1] suggests the peculiarity of the electronic structure of MgB$_2$. Observation of the isotope effect[2] suggested the importance of the electron-phonon coupling in MgB$_2$. The band structure of MgB$_2$ has been energetically investigated[3-9], and the bands near the Fermi level are two σ-bands which are mainly formed by boron p$_{x,y}$ orbitals, and two π-bands which are mainly formed by boron p$_z$ orbitals. A strong electron-phonon coupling with the dimensionless coupling constant $\lambda \geq 0.7$ was suggested[4-6,8] mainly due to the large electron-ion matrix element, especially for the σ-bands. However, this value may be overestimated due to, e.g., the rigid muffin-tin approximation, which is justified only for not heavily anisotropic (namely cubic) lattices[10]. Specific heat measurements[11-13] showed the enhancement factor $1+\lambda+\lambda_{e-e} \simeq 1.6$ (here $\lambda_{e-e}$ stands for the electron-electron coupling constant), hence gives the upper limit of $\lambda$ as 0.6. In any case, $T_c$=40K seems too high to be attributed solely to the large $\lambda$, and there seems still room for other mechanisms. Given the McMillan formula[14], the intermediate value of $\lambda$ and high $T_c$=40K formally gives a very small, or even negative value of the Coulomb pseudo-potential term $\mu^*$. The possibility of negative or small $\mu^*$ is recently discussed in the framework of two-band mechanism[15]. This model naturally gives a small value, or even negative value of $\mu^*$ based on an assumption that interband polarization function $\Pi_{12}(\mathbf{Q})$ is enhanced for $\mathbf{Q}\sim(0,0,\pi/c)$. This scheme leads to a multi-gap superconductivity with larger gaps in the π-bands and subsidiary small gaps in the σ-bands. In accordance with this framework, observed multi-gap features of specific heat data[12,13] and photoemission data[16] both suggest the multi-gap nature of superconductivity. Specific heat experiments[13] indicate further that the smaller superconducting gap (~10K) is associated with a band or bands with the state density slightly larger than 30% of the total state density. According to the band calculation, this number is consistent with the above-mentioned two band picture which requires the small gap to belong to the σ-bands. On the other hand, photoemission data[16] indicate that the state with smaller (=1.7meV) gap has an intensity five times larger than the state with larger (=5.6meV) gap. Thus the experimental determination whether the wider gap belongs to the σ-band or the π-band awaits the success of growth of the single crystal. Even though, the photoemission result is also interpretable in the same picture if one takes account of the extent of the wave function, as will be discussed later. Thus the dominant role played by the π-bands for the superconductivity is considered to be supported by the availeable experimental data. Therefore, it is essentially important to investigate the growth of interband polarization function $\Pi_{12}(\mathbf{Q})$ for $\mathbf{Q} \sim (0,0,\pi/c)$.

However, the author of this theory[15] estimated $\Pi_{12}(\mathbf{Q})$ only semiquantitatively. In his estimation the whole shape of the π-bands are not correctly treated. He employed a linear approximation near the doubly degenerate points on the K-H axis in the Brillouin zone, and suggested its enhancement for the wave vector linking the two degeneracy points. In this paper we have performed a numerical calculation of $\Pi_{12}(\mathbf{Q})$ and its temperature dependence on the basis of an *ab-initio* band calculation, and found that $\Pi_{12}(\mathbf{Q})$ becomes quite large for $\mathbf{Q} \sim (0,0,Q_z)$ with $Q_z$ lying in the region around π/c as expected from the closeness of the Fermi surface to perfect nesting[17]. Further we obtained a richer structure of $\Pi_{12}(\mathbf{Q})$, i.e., two peaks of $\Pi_{12}(\mathbf{Q})$; one of the peaks is near but appreciably shifted shifted from what we expect from the tight-binding band, and the other peak located at $Q_z=\pi/c$ is an unexpected one. This result has an important meaning in an exotic mechanism of superconductivity[17], bringing in the negative or smaller μ*. We also observed an isotope effect and a pressure effect in the calculated $\Pi_{12}(\mathbf{Q})$, due to the change of the electronic band structure accompanying the expansion or contraction of the lattice. These depenences are qualitatively in accordance with the experimentally observed shift of $T_c$[2,18] according to the two-band mechanism[15].

We performed an *ab-initio* band calculation based on the full-potential linearized augmented plane wave (FLAPW) method by using the computer code KANSAI-94. We carried out the calculation for the three lattice parameter sets shown in Table I for investigating the observed isotope[2] and pressure effect[18]. Hereafter we call the sample of $MgB_2$ with naturally abundant B as $Mg^{11}B_2$, in order to distinguish it from $Mg^{10}B_2$. For the one-electron-exchange-correlation potential we use the local-density approximation (LDA) scheme according to the prescription of Gunnarson and Lundqvist[18]. The muffin-tin radius of each atom is r(Mg)=1.48Å and r(B)=0.90Å for all compounds. The calculation of the core states and the valence states are self-consistently carried out by the scalar-relativistic scheme. We used the basis functions with the wave vector $|\mathbf{k}+\mathbf{G}| < K_{\max} = 3.60(2\pi/a)$, where $\mathbf{k}$ is a wave vector in the Brillouin zone, and $\mathbf{G}$ is a reciprocal-lattice vector resulting in about 300 basis LAPWs. The self-consistent potentials was calculated at 40 points (or 21 points, but the result was almost unchanged) in the irreducible Brillouin zone (IBZ). The density of states (DOS) was deduced from the final eigenstates calculated at 549 points. Total energy calculation reveals that $Mg^{11}B_2$ ($a=a_0,c=c_0$) is the most stable among these three compounds at zero pressure.

The obtained energy bands are quantitatively consistent with previous works[4-9]. The states near the Fermi energy ($E_F$) almost come entirely from the B-p states. We show the DOS at $E_F$ ($D(E_F)_{\text{calc}}$) for the three sets of lattice constants in Table I. The DOS at $E_F$ is almost unchanged between Mg$^{11}$B$_2$ and Mg$^{10}$B$_2$, and decreases for the compressed Mg$^{11}$B$_2$(press.). $D(E_F)_{calc}$ as a function of the lattice constant $a$ takes a positive slope as expected from a single-band model, but its non-linearity is unexpected. The muffin-tin projected DOS at the boron site shows the partial DOS of the boron p$_{x,y}$ bands is larger than that of the p$_z$ bands, which apparently disagrees with state densities of σ- and π-bands reported by Liu et al.[9], but this discrepancy is understandable because the π-band states are actually more extended outside of the boron muffin-tin spheres than the σ-band states, having a larger Mg contribution. This result allows a reasonable interpretation resolving for the apparent contradiction between the specific heat data[13] and photoemission data[16], if we assume that the larger gap belongs to the π-bands. Since photoemission measurements with photon energy $E_{ph}$=21.2eV mainly probes the B-2s and B-2p states, the state with stronger intensity should be the σ-bands. The σ-bands give a larger intensity to the states with small gap due to a heavier B-p component, whereas the π-bands give a smaller intensity for the states with the large gap. On the other hand since specific heat measurement probes the total DOS, therefore the small gap should be ascribed to the bands with smaller DOS, namely the σ-bands. Thus both kinds of data are consistently interpretable with the picture that the σ- and π-bands carry the small and large gaps, respectively.

Three Fermi surfaces (from the 3rd to the 5th band) are obtained in this calculation. We show the results only for Mg$^{11}$B$_2$. The 3rd band forms a cylindrical hole Fermi surface (FS) which comes from the B(sp)$^2$ orbitals (σ-band). The 4th band forms two kinds of FS's. One is the cylindrical σ-band hole FS around the Γ-A axis which resembles to that in the 3rd band, and another hole FS which forms a honeycomb-like network in the $k$-space, and it comes from B-p$_z$ orbitals (π-band). At general $k$-points in the 4th bands the σ- and the π-band are mixed and a crossover of band nature occurs in the **k**-space. Finally the 5th band forms an elecron FS whose shape is nearly a mirror image of the π-band portion of the FS in the 4th band. This band also comes from the π-band. We show energy contour plots of the 4th and the 5th band in the ($k_x$,$k_z$)- and ($k_y$,$k_z$)- section planes in Fig. 1. These two bands have the indentical energy along the K-H axis (vertical dotted line), this double degeneracy is lifted elsewhere than along this axis. The band energy along the axis crosses the Fermi energy at the P point where KP~5/12 KH. In the neighborhood of P the hole-type FS in the 4th band has a shape of circular cone. In the 5th band there is an electron FS which has almost a mirror image shape with respect to a plane parallel to the ($k_y$,$k_z$)-plane and crossing HK line at P. If we move the electron FS downwards by twice the distance KP, it nests with the hole FS in the neighborhood of the P' points, which is the mirror image of P with respect to the ($k_x$,$k_y$)-plane. Thus an enhancement of $\Pi_{12}(\mathbf{Q})$ is expected for $Q_x=Q_y=0$ and $Q_z$ equal to the distance between P and P'. Furukawa[17] made another intriguing remark that if the σ-band hole pockets did not appear the electron FS should have a complete nesting when the bands are correctly described by the bipartite tight-binding band.

The interband polarization function $\Pi_{12}(\mathbf{Q})$ is defined by

$$\Pi_{12}(\mathbf{Q}) = \frac{1}{N} \sum_{\mathbf{k}} \frac{f(\varepsilon_1(\mathbf{k+Q})) - f(\varepsilon_2(\mathbf{k}))}{\varepsilon_2(\mathbf{k}) - \varepsilon_1(\mathbf{k+Q})}$$

where $N$ is the number of unit cells and $f(\varepsilon)$ is the Fermi-Dirac distribution function. If we regard $\varepsilon_{1,2}(\mathbf{k})$ as the calculated 4th and the 5th bands, this quantity is directly calculated by the obtained results of the bands $E(\mathbf{k})$. We show the results of **Q**- and *T*- dependences of $\Pi_{12}$ in Fig. 2. The most important feature is the growth of $\Pi_{12}(\mathbf{Q})$ in the range of $0.25(2\pi/c) < Q_z \leq 0.5(2\pi/c)$. This is considered to come from the closeness of FS's to the perfect nesting. Secondly, two peak structures at $Q_z=Q_1 \sim 0.3(2\pi/c)$ and $Q_z=Q_2=0.5(2\pi/c)$ are clearly seen, and its $T$-dependence resembles the analytic result[15] in a broad sense. However, the peak position of $Q_1$ is not precisely equal to the above-mentioned distance PP' which gives $Q_z=Q_0 \sim 0.4(2\pi/c)$. We attribute this to the fact that the nesting along the M-L line is better in a wider region. One can confirm this by reversing the 4th band (Fig. 1(a)), shifting it along $k_z$-axis and then placing it upon the 5th band (Fig. 1(b)), then both the left and right panels nests very well. This nesting is not due to the double degeneracy but rather accidental one. The absolute value of $\Pi_{12}(Q)=0.27[eV^{-1}]$ at $Q_z=Q_2$ is remarkably large in view of the state densities, ~0.093 and ~0.112 $[eV^{-1}]$, of the two π-bands, respectively[9]. It supports the two-band framework in ref.15 that the effective interband pair-transfer coupling constant $\widehat{K} \equiv K/[1-(K+U')\Pi_{12}(Q)]$ grows divergently and drives two-band superconductivity if $(K+U') \equiv U_0$ approaches 3.6eV, where $K$ denotes the interband electron-pair scattering and $U'$ denotes the intraband electron-pair scattering. Even if $\widehat{K}$ does not diverge, the large value of $\widehat{K}$ leads to the small or negative μ*. We note that this enhancement of $\Pi_{12}(Q)$ also appears for the **Q** vector not strictly equal to $(0,0,Q_z)$. When the vector **Q** takes the form of $(Q_x,Q_y,Q_z)$ where $Q_x$ and $Q_y$ are small quantities (namely ~ 1/24a*), $\Pi_{12}(Q)$ decreases because of the deterioration of the nesting, but its decrease is about 5% in the case of $Mg^{11}B_2$. Thus the remarkable enhancement of $\Pi_{12}(Q)$ occurs in a considerably large part in the **Q**-space. We also note that other combinations of bands, for example σ- and π-bands, do not give such a large value of $\Pi_{12}(Q)$. The calculated $\Pi_{12}(Q)$ between the 3rd band (=σ-band) and the 5th band (=π-band) is almost $Q_z$-independent and its value is about $0.15[eV^{-1}]$. This result is naturally understood since the energy of the σ-band is almost $k_z$-independent, as clearly seen in Fig. 1(a).

When the two boron atoms in the unit cell are located at $(\pm a/2\sqrt{3},0,0)$, then the tight-binding bands of the boron $p_z$-orbitals are given by [20]

$$\varepsilon_{TB}^{\pm} = \pm t\left\{3 + 2\cos(ak_y) + 4\cos\left(\frac{\sqrt{3}}{2}ak_x\right)\cos\left(\frac{1}{2}ak_y\right)\right\} + 2t_z\cos(ck_z) + \varepsilon_0$$

The band parameters are given by Kortus et al.[4] as $t \sim 2.5$eV and $t_z \sim 1.5$eV, which also give a good approximation of the presently calculated FLAPW bands. The offset value $\varepsilon_0=0.4474$Ry is chosen so as to fit the FLAPW $\pi$-bands well especially for the valence band. We show the energy contour plots of these bands in Fig. 3. We can see that if chemical potential $\mu=0$ (symmetric case), these two bands $\varepsilon_{TB}^{\pm}$ perfectly nest and presumably cause an antiferromagnetic instability[17]. However, the $\sigma$-band deprives the $\pi$-bands of its holes and the FS's become asymmetric. In this case the FS of $\varepsilon^+$ expands and the FS of $\varepsilon^-$ shrinks, then the above-mentioned vectors $Q_1$ and $Q_2$ do not remain nesting vectors, and only the doubly degenerated P and P' points can form nesting vector $Q_0$. However fortunately, the nesting among the FLAPW bands is better than the expected from the tight-binding band. We schematically show this situation in Fig. 4. The band dispersion $E(\mathbf{k})$ of the FLAPW band is not strictly fitted by the above tight-binding bands especially for the electron band due to the hybridization with the Mg sp-band. Then rather accidental nesting vectors $Q_1$ and $Q_2$ brings about the peaks of $\Pi_{12}(Q_z)$; here $\Pi_{12}(\mathbf{Q})$ with $Q=(0,0,Q_z)$ is abbriviated to $\Pi_{12}(Q_z)$.

Finally we comment on the isotope and the pressure effects in $MgB_2$. We show $\Pi_{12}(Q_z)$ for various lattice constants (see Table I) in Fig. 5, with shifting Fermi energy mimicing electron and hole doping. First we examine the solid curves which are for the non-doped cases. In the case of $Mg^{10}B_2$ (the lattice constant $a$ is expanded by 2% relative to $Mg^{11}B_2$), the peak at $Q_2$ overwhelms the peak at $Q_1$. The absolute value of $\Pi_{12}(Q_z)$ increases relative to Mg$^{11}$B$_2$. In the case of Mg$^{11}$B$_2$(press.), Q$_2$ peak is lower than Q$_1$ peak, and the absolute value of these peaks are considerably smaller. Therefore, expansion of the lattice should raise $T_c$ and contraction of the lattice should suppress $T_c$ according to the scheme of ref.15. These findings are consistent with the isotope effect[2] and the pressure effect[18]. We note that D($E_F$) changes only little between the three lattice parameter sets (see Table I), indicating that D($E_F$) is not so sensitive to the shape of the FS. On the other hand, the quantity $\Pi_{12}(Q_z)$ is very sensitive to the shape of the FS's. Next we examine the doping dependence of $\Pi_{12}(Q_z)$ for each compound. As for the electron-doped case, $\Pi_{12}(Q_z)$ decreases in all compounds. This is because of the deterioration of the nesting due to the aggravated dissimilarity of the electron and the hole FS's due to electron doping. On the other hand, hole doping always increases $\Pi_{12}(Q_z)$, and a downward shift by about 30mRy of $E_F$ significantly increases $\Pi_{12}(Q_z)$. This shift allows $E_F$ to reach the energy at the midpoint of the K-H axis, hence brings about the symmetric case in the tight-binding model. In this case $\Pi_{12}(Q_z)$ has only one large peak at Q$_z$=0.5(2$\pi$/c) as expected from the tight-binding model. However, the nesting is not perfect in the real band dispersion and the expected log$T$ divergence is absent. Even though, we expect that increase of the $\Pi_{12}(Q_z)$ brings about a large increase of $\widehat{K}$, therefore a significant decrease of $\mu^*$ and thus a large increase of $T_c$ by introducing ~0.3 holes in the formula unit, if the rigid band model holds and the nesting feature stays basically unchanged by doping. However in this case, a competition between superconductivity and other instabilities such as antiferromagnetism[17] and CDW along the *c*-axis may take place. This may be a cause of suppression of superconductivity in heavily hole-doped system Mg$_{1-x}$Li$_x$B$_2$[21,22].

In summary, we have calculated the band structure of MgB$_2$ for three sets of lattice constants and found substantial increase of the interband polarization function between two π-bands due to considerable inter π-band nesting of appropriate wave vectors along the $k_z$-axis. This behavior is in accord with the general expectation, but a close examination reveals a richer way of enhancement than expected from the tight-binding model built of boron $p_z$-orbitals, suggesting the importance of the hybridization with Mg sp-bands. The obtained enhanced polarization function is probable to cause a superconducting instability in $MgB_2$ or at least a decrease of the Coulomb pseudo-potential term $\mu^*$, according to the two-band mechanism. Hole doping increases the polarization function and thus expected to lead to a the further increase of $T_c$. The increase of $T_c$ due to isotopic replacement of $^{11}B$ by $^{10}B$ and the decrease of $T_c$ under hydrostatic pressure were ascribed to the increase and decrease of the same function, respectively, through band modification. An anomalous increase of the interband spin susceptibility of similar nature should be observable in neutron scatering experiments.

We thank to T. Yanagisawa and T. Yokoya for useful discussions. Numerical computation was mainly performed at Tsukuba Advanced Computing Center at National Institute of Advanced Industrial Science and Technology.

# References

1) J. Nagamatsu, N. Nakagawa, T. Muraoka, Y. Zenitani and J. Akimitsu: Nature **410** (2001) 63.

2) S. L. Bud'ko *et al.*: Phys. Rev. Lett. **86** (2001) 1877.

3) D. R. Armstrong, and P. G. Perkins: J. C. S. Faraday II, **75** (1979) 12.

4) J. Kortus *et al.*: cond-mat/0101446.

5) J. M. An and W. E. Pickett: cond-mat/0102391 (2001).

6) Y. Kong *et al.*: cond-mat/0102499 (2001).

7) N. I. Medvedeva *et al.*: cond-mat/0103157 (2001).

8) T. Yildirim *et al.*: cond-mat/0103469 (2001).

9) A. Y. Liu *et al.*: cond-mat/0103570 (2001).

10) G. Gaspari and B. Gyorffy: Phys. Rev. Lett. **28** (1972) 801.

11) Y. Wang *et al.*: cond-mat/0103181 (2001).

12) Ch. Wälti *et al.*: cond-mat/0102522 (2001).

13) F. Bouquet i. cond-mat/0104206 (2001).

14) W.L. McMillan: Phys. Rev. **167** (1968) 331.

15) K. Yamaji: cond-mat/0103431 (2001), to be published in J. Phys. Soc. Jpn. **70**(6).

16) S. Tsuda *et al*. cond-mat/0104489 (2001).

17) N. Furukawa: cond-mat/0103184 (2001).

18) K. Prassides *et al*.: cond-mat/0102507; T. Vogt *et al*.: cond-mat/0102480; J. D. Jorgensen *et al*.: cond-mat/0103069; T. Tomita *et al*.: cond-mat/0103538 (2001).

19) O. Gunnarson and B. I. Lundqvist: Phys. Rev. B **13** (1976) 4274 .

20) P.R. Wallace: Phys. Rev. **71** (1947) 622 .

21) Y. G. Zhao *et al*.: cond-mat/0103077 (2001).

22) J. Y. Xiang *et al*.: cond-mat/0104366 (2001).

Figure Captions

Fig. 1. The ($k_x$,$k_z$) and ($k_y$,$k_z$) section planes of energy contour of (a) the 4th band and (b) the 5th band of Mg$^{11}$B$_2$. Units of the axes are the length of reciprocal lattice vectors in each direction, when we take body-centered orthorhombic lattice as the unit cell. The unit of the energy is Ry (1Ry=13.598eV). The solid curve is the Fermi surface ($E$=$E_F$=0.4887Ry).

Fig. 2. Calculated polarization function $\Pi_{12}(Q_z)$ of Mg$^{11}$B$_2$ for various temperatures.

Fig. 3. The ($k_x$,$k_z$) and ($k_y$,$k_z$) section plane of energy contour of tight-binding (a) $\varepsilon^+$ band and (b) $\varepsilon^-$ band. The dot-dashed and the solid lines indicate the Fermi surfaces for symmetric case ($\mu$=0, $E_F$=0.4474Ry) and asymmetric case ($\mu$=0.0413, $E_F$=0.4887Ry), respectively.

Fig. 4. A schematic figure of Fermi surface nesting in MgB$_2$ based on the FLAPW calculation. Arrows denote nesting vectors $Q_z$= $Q_0$~0.4($2\pi$/c), $Q_z$=$Q_1$~0.3($2\pi$/c) and $Q_z$=$Q_2$=0.5($2\pi$/c). The solid curve denotes the FS of the real band. The dotted curves and dot-dashed curve denote the tight-binding bands for symmetric case ($\mu$=0, $E_F$=0.4474Ry) and asymmetric case ($\mu$=0.0413Ry, $E_F$=0.4887Ry), respectively.

Fig. 5. Calculated polarization function $\Pi_{12}(Q_z)$ of Mg$^{10}$B$_2$, Mg$^{11}$B$_2$ and Mg$^{11}$B$_2$ (press.) in doped situations. Solid curves are for the case with no $E_F$ shift, and other curves are the case with shifting $E_F$ by -30mRy,

-10mRy (hole-doping) and +10mRy (electron-doping). Temperature $T$ is fixed to 158K.

| material | lattice constants [Å] | | $D(E_F)_{calc}$ [States/(eV f.u.)] | | |
| --- | --- | --- | --- | --- | --- |
| | $a$ | $c$ | total | B-$p_{x,y}$ | B-$p_z$ |
| Mg$^{10}$B$_2$ | 3.1432[a] (~1.02$a_0$) | 3.5193[a] (~$c_0$) | 0.729 | 0.161 | 0.110 |
| Mg$^{11}$B$_2$ | 3.083 (=$a_0$) | 3.521 (=$c_0$) | 0.724 | 0.160 | 0.104 |
| Mg$^{11}$B$_2$(press.) | 3.021 (=0.98$a_0$) | 3.521 (=$c_0$) | 0.696 | 0.157 | 0.096 |

[a] ref.2.

Table I. The lattice parameters of MgB$_2$ used in this calculation, and the calculated density of states. The lowest column Mg$^{11}$B$_2$(press) denotes Mg$^{11}$B$_2$ under high pressure. The pressure resulting in this lattice parameter is not calculated.

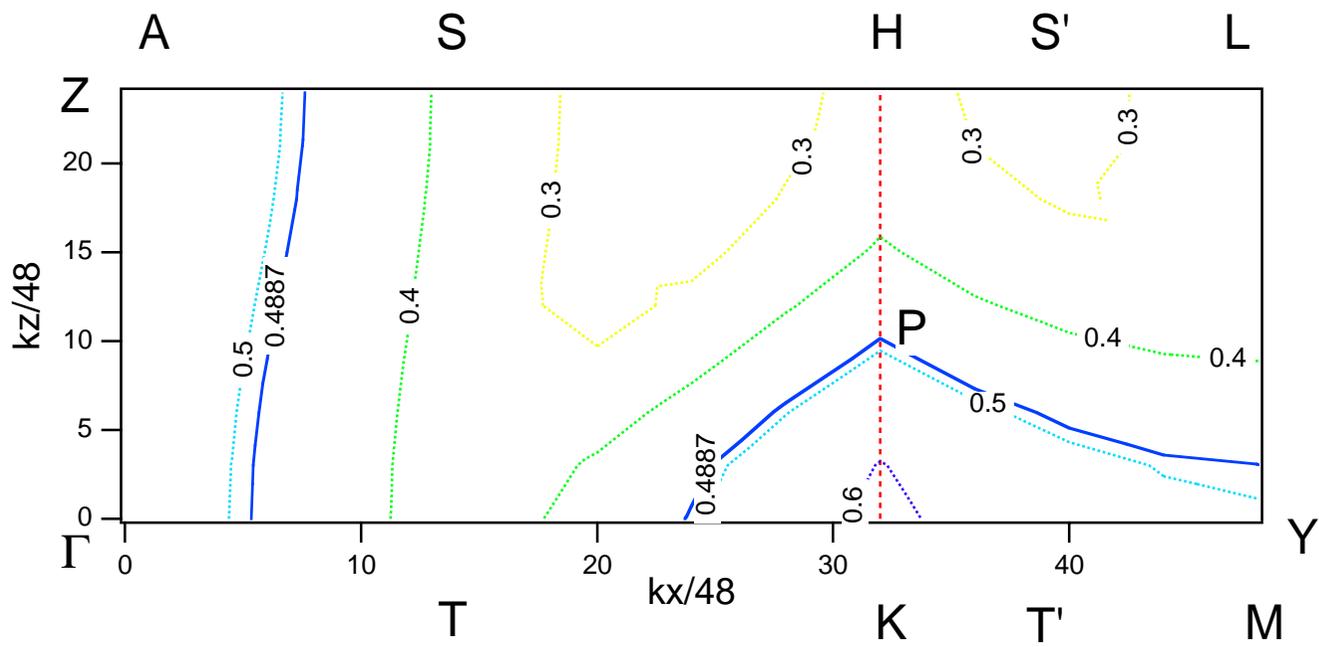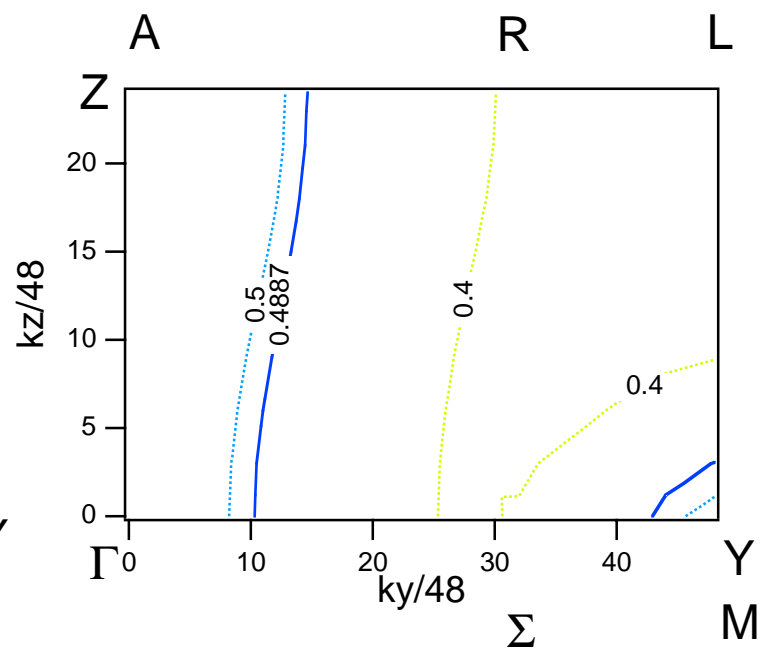

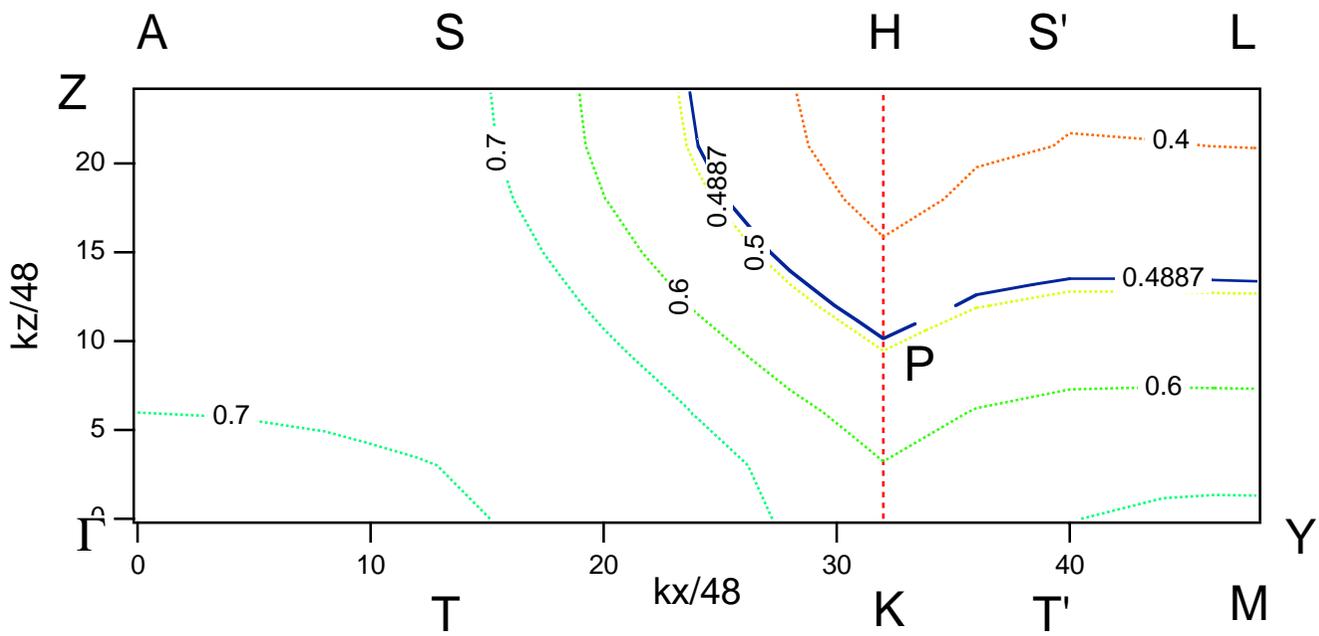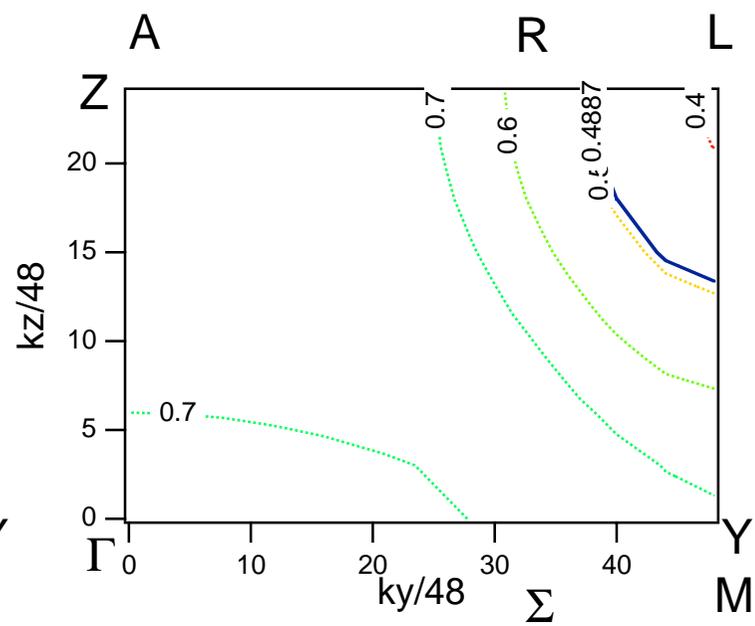

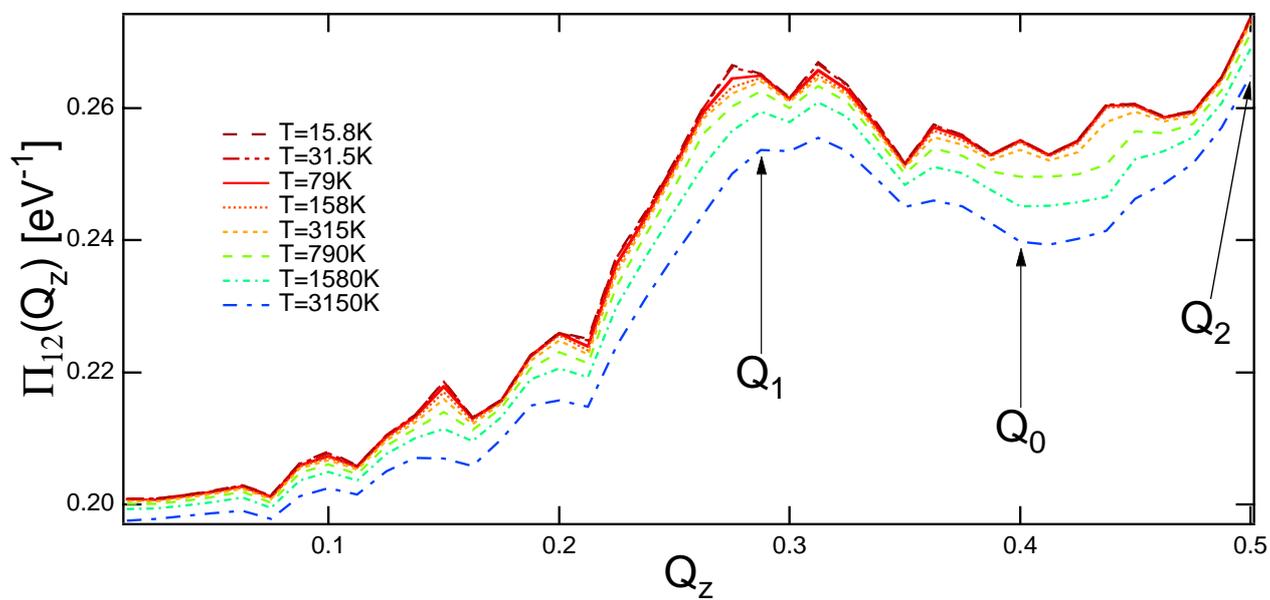
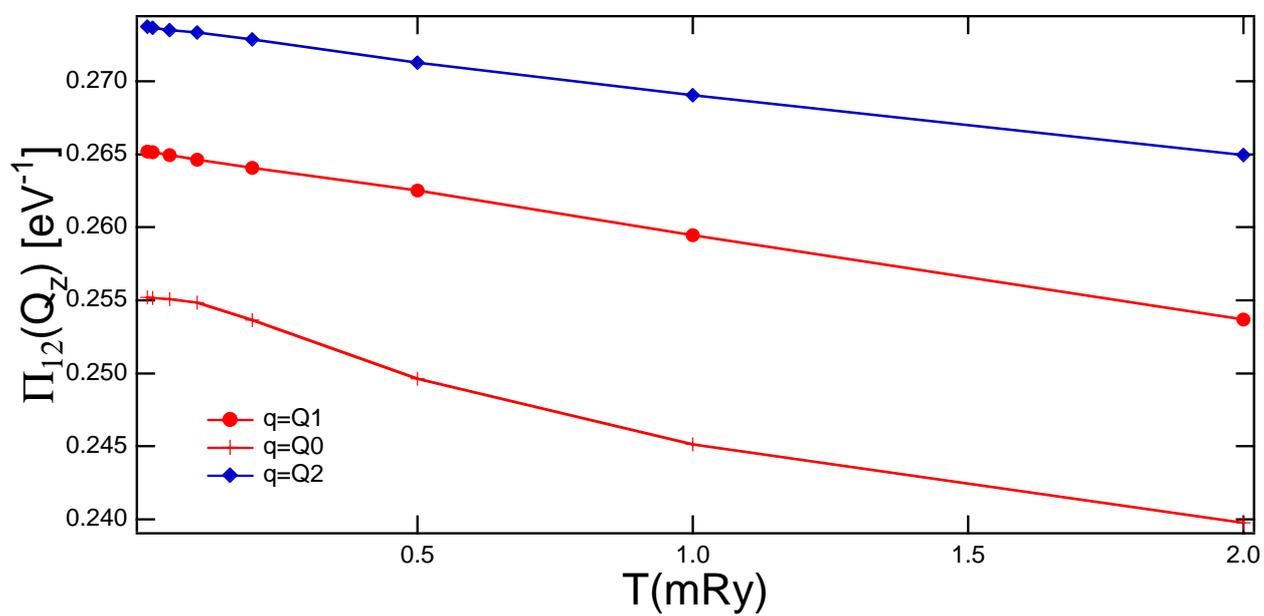

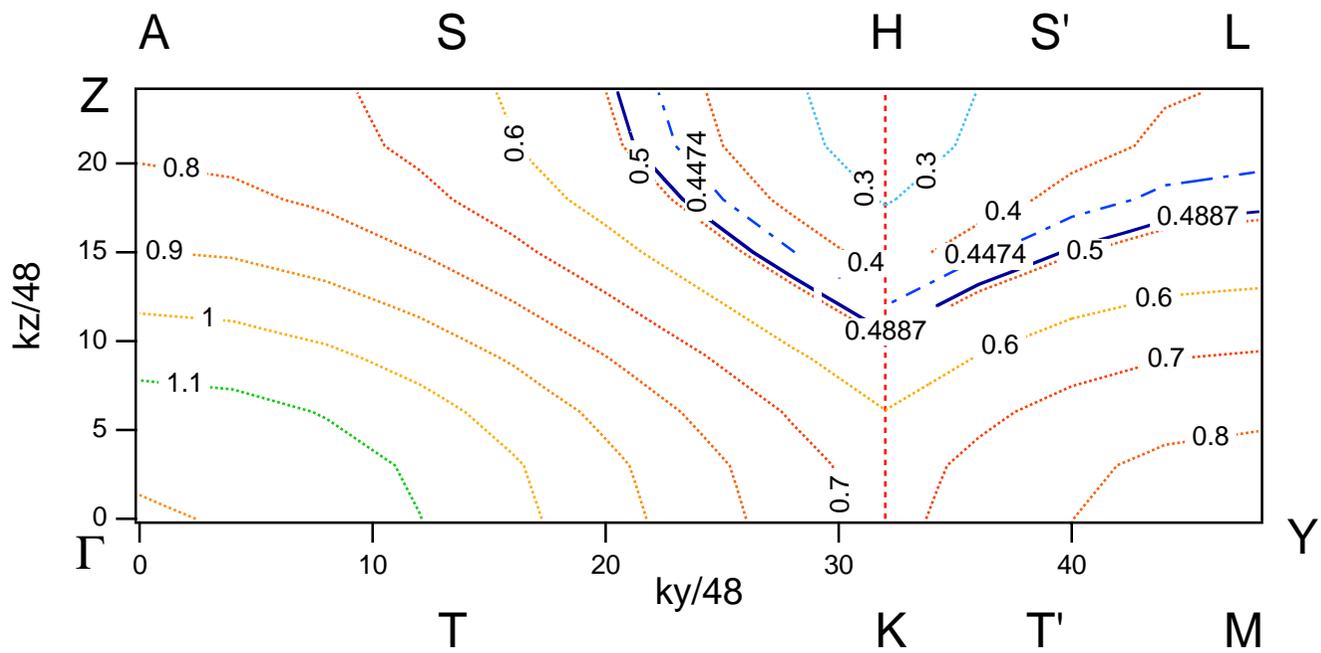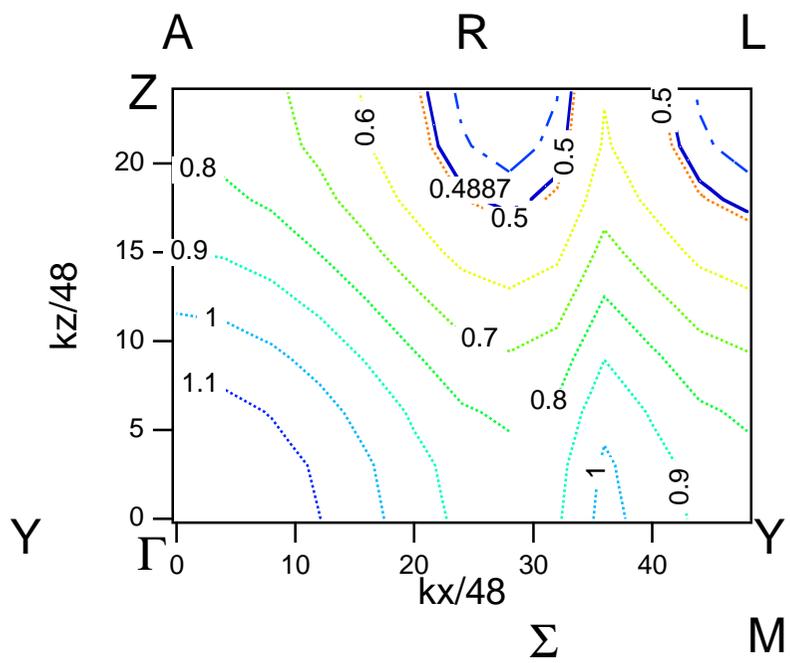

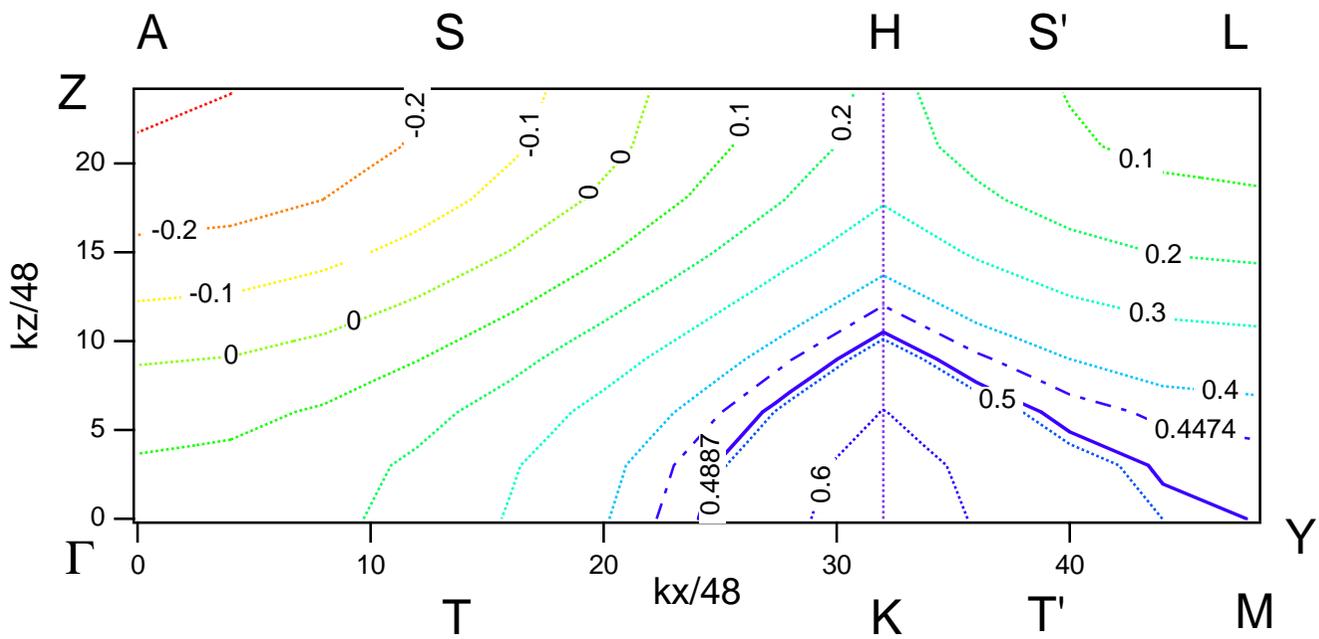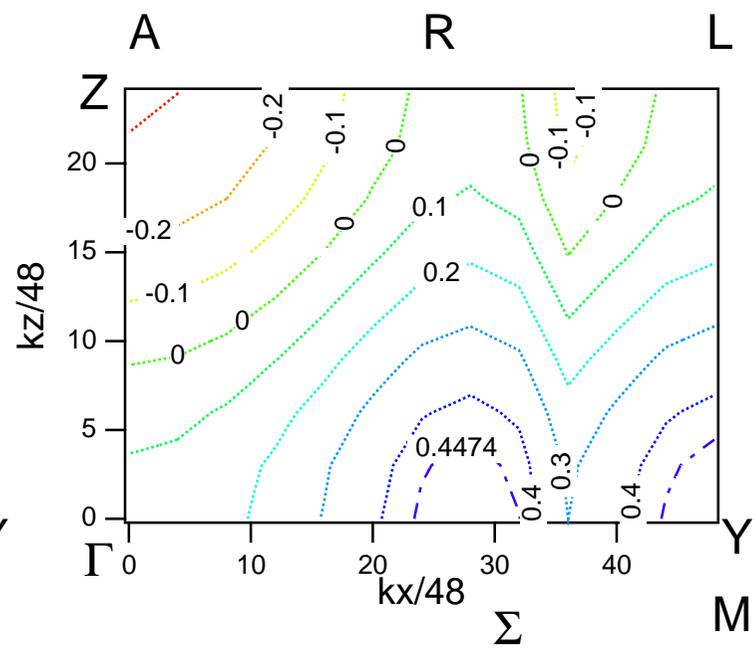

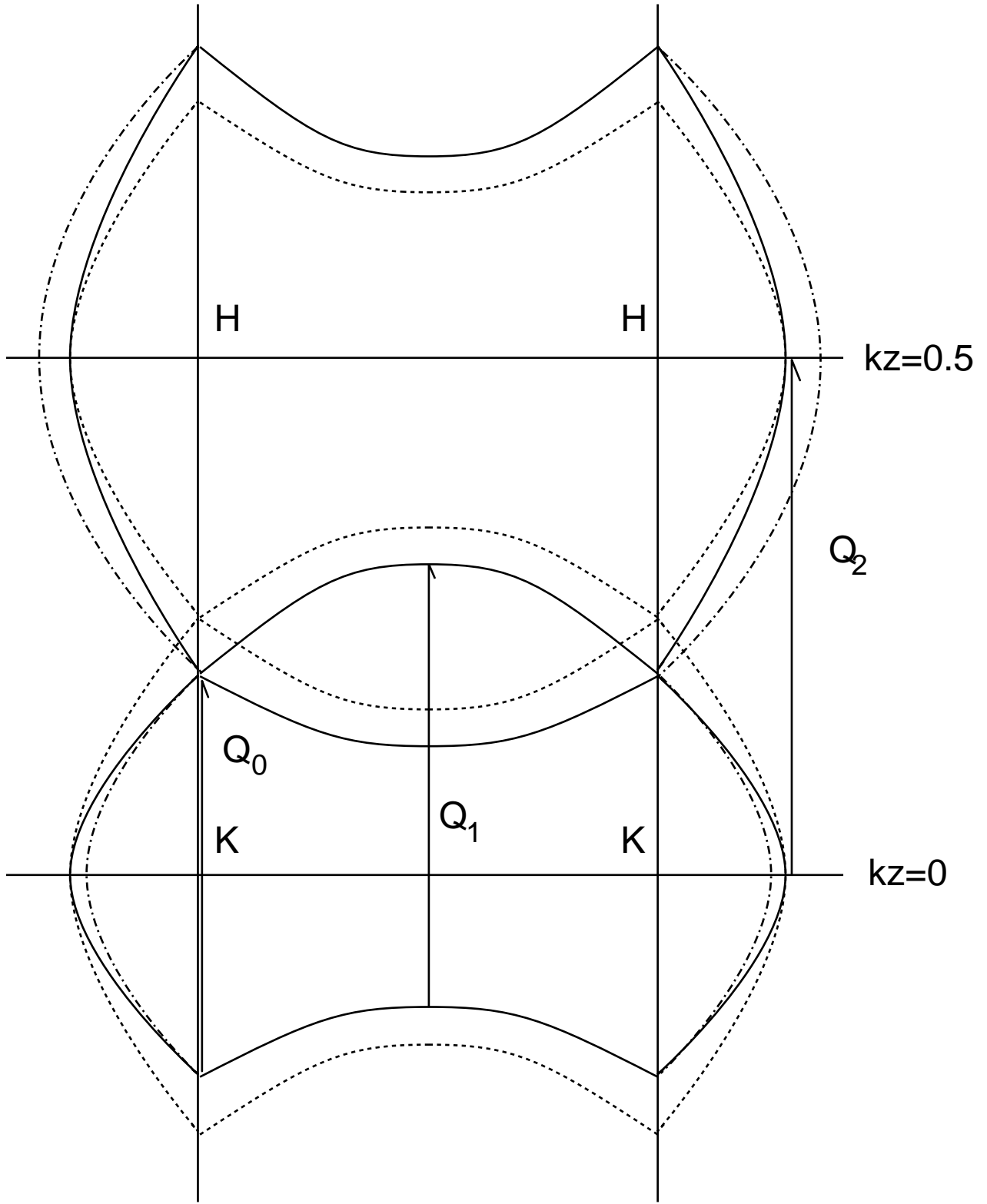

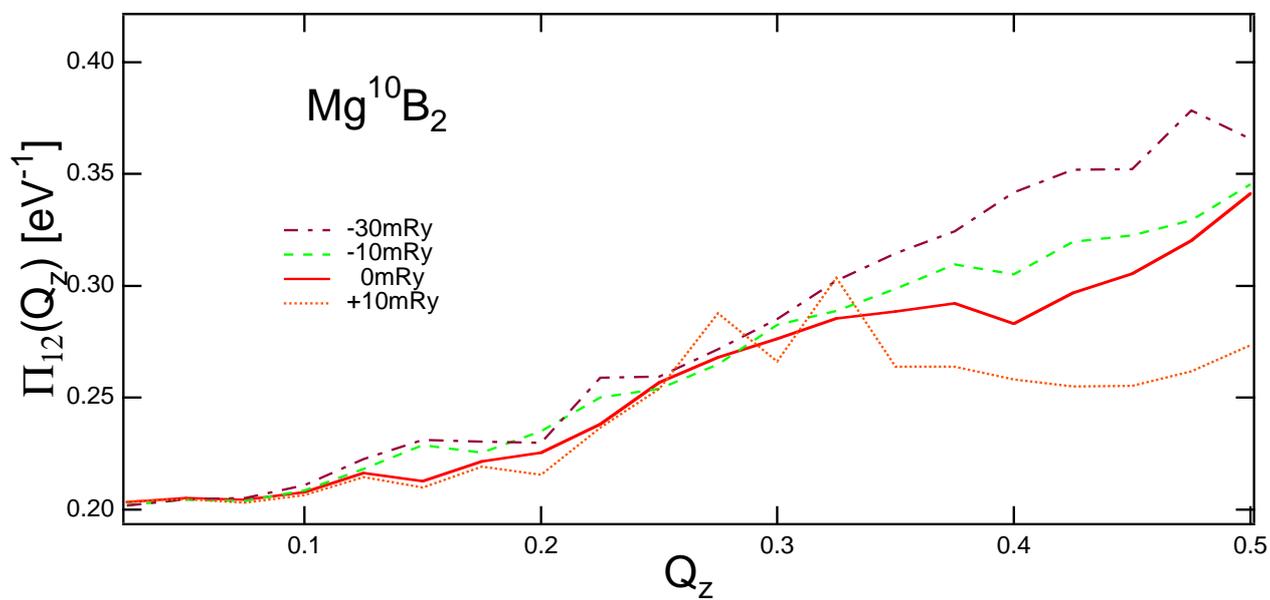
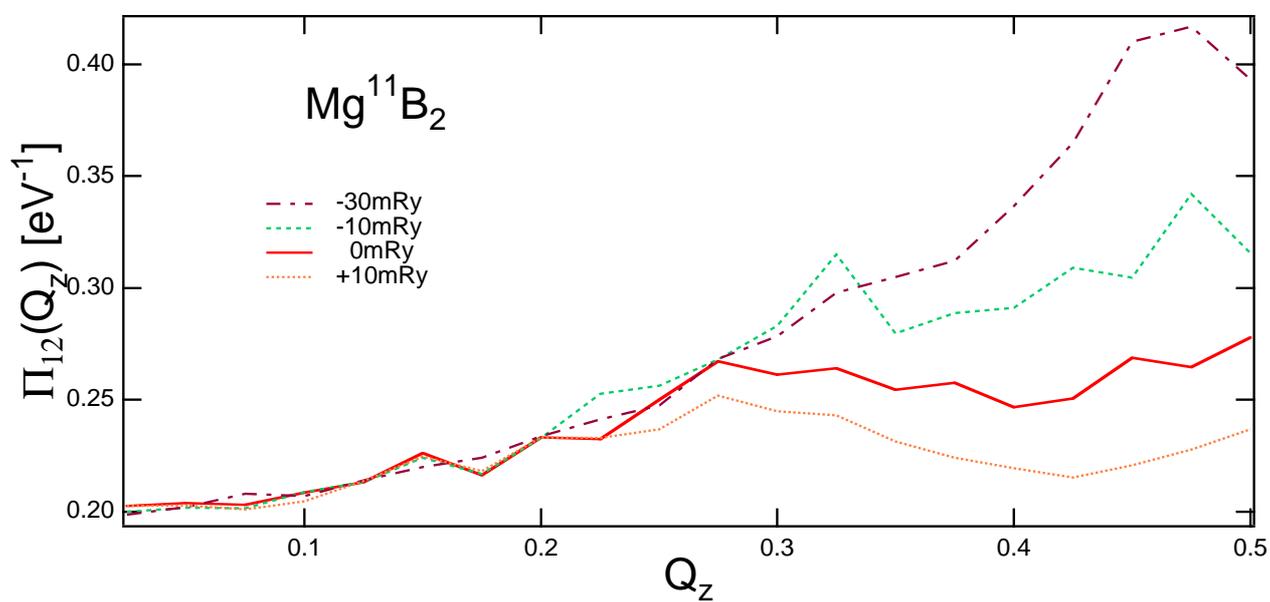
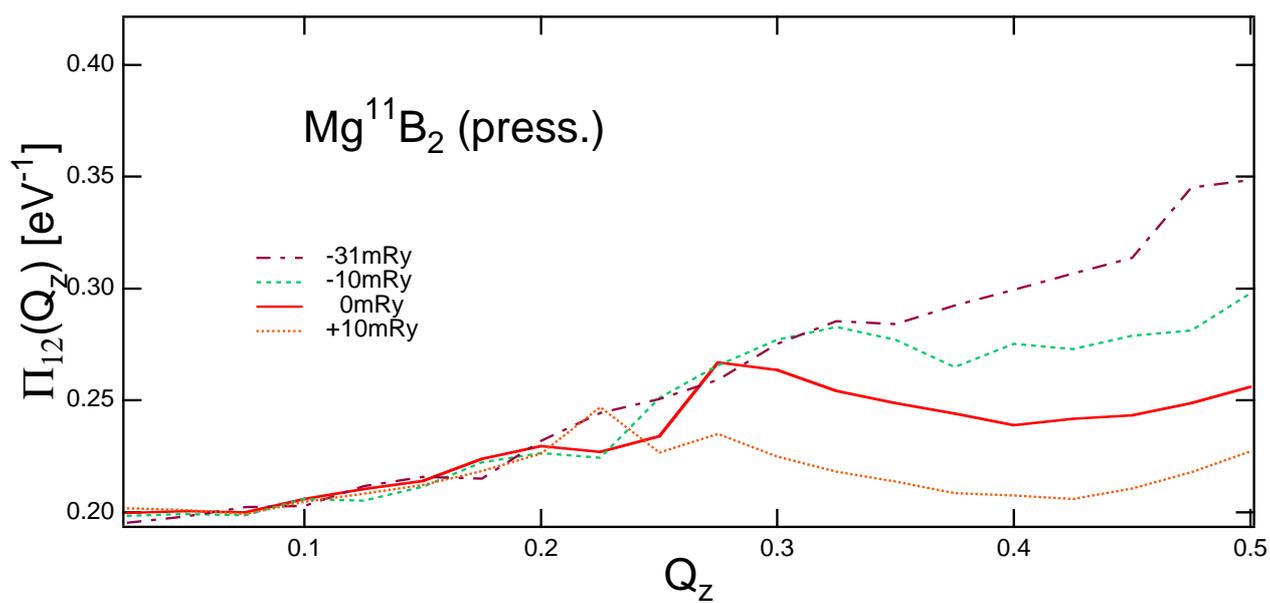